\documentclass[prl,twocolumn,showpacs]{revtex4-1}
\usepackage[margin=1in]{geometry}
\usepackage{hyperref}
\hypersetup{
  colorlinks   = true, 
  urlcolor     = blue, 
  linkcolor    = blue, 
  citecolor   = red 
}
\usepackage{graphicx}
\usepackage{amsmath}
\renewcommand{\vec}{\mathbf}

\usepackage{color}

\begin{document}

\title{Antiferromagnetic Domain Wall Motion Induced by Spin Waves}

\author{Erlend G. Tveten}
\author{Alireza Qaiumzadeh}
\affiliation{Department of Physics, Norwegian University of Science and Technology, NO-7491 Trondheim, Norway}
\author{Arne Brataas}
\affiliation{Department of Physics, Norwegian University of Science and Technology, NO-7491 Trondheim, Norway}
\date{\today}

\begin{abstract}
Spin waves in antiferromagnets are linearly or circularly polarized. Depending on the polarization, traversing spin waves alter the staggered field in a qualitatively different way. We calculate the drift velocity of a moving domain wall as a result of spin wave-mediated forces and show that the domain wall moves in opposite directions for linearly and circularly polarized waves. The analytical results agree with micromagnetic simulations of an antiferromagnetic domain wall driven by a localized, alternating magnetic field.
\end{abstract}


\maketitle
Antiferromagnets (AFMs) are promising candidates for future spintronic devices for the following reasons: 1) They can be integrated with ferromagnetic components; 2) switching occurs at very high frequencies; and 3) there are no stray fields, allowing small independent devices to be created ~\cite{MacDonald13082011, Duine:2011fk}. The dynamics of AFMs are fundamentally different from those of ferromagnets (FMs) because the equations of motion are second order in frequency rather than first order~\cite{PhysRevLett.110.127208, PhysRevLett.106.107206}. Furthermore, AFMs are affected by both charge and spin currents, as was recently shown theoretically ~\cite{ PhysRevB.73.214426,*PhysRevB.75.014433, *PhysRevB.75.174428, *PhysRevLett.100.196801, *PhysRevB.81.144427,*PhysRevB.85.134446,*PhysRevB.89.081105} and experimentally~\cite{PhysRevLett.98.116603, *PhysRevLett.98.117206, *PhysRevLett.99.046602}. The antiferromagnetic order can be probed, e.g., via the anisotropic tunneling magnetoresistance effect ~\cite{Park:2011rt,*Marti:2014xy,*PhysRevLett.108.017201,*PhysRevB.79.134423,*PhysRevB.81.212409}. Additionally, a change in the spin texture of the AFM affects both the longitudinal and Hall resistivities~\cite{Kummamuru:2008fk,*Soh28092011}.

In AFMs, domains usually result from crystal imperfections~\cite{PhysRev.126.78}, but they may also inherit the domain structure of the ferrimagnetic precursor layer as they undergo a phase transition to the antiferromagnetic phase~\cite{PhysRevLett.106.107201}. Antiferromagnetic domains~\cite{Nolting:2000uq} and several forms of domain wall (DW) structures in AFMs have been observed~\cite{Bode:2006uq}. Furthermore, DWs in AFMs can also be induced, controlled and engineered by exchange bias pinning forces~\cite{logan:192405,PhysRevLett.110.127208}. 

Progress in the field of antiferromagnetic spintronics requires the development of novel methods for exciting AFMs at the nanoscale. Many AFMs are insulating and cannot be affected by currents in the bulk; however, other approaches can be employed to excite an AFM. We suggest the use of antiferromagnetic spin waves (SWs) as a new and exciting way of manipulating the order of AFMs. The advantage to this method is that SWs in AFMs operate coherently in the THz regime~\cite{Kampfrath:2011uq}, which is orders of magnitude faster than the frequency of typical ferromagnetic SWs.

In this Letter, we demonstrate that SWs move DWs in AFMs. We show that this phenomenon is considerably richer than the analogous SW-DW interaction in FMs due to the inherent complexity of antiferromagnetic SWs~\cite{:/content/aapt/journal/ajp/21/4/10.1119/1.1933416}. In contrast to SW-driven DW motion in FMs, we find that the direction of DW motion in AFMs is governed by the nature of the SW excitation modes. This behavior enables superior control of DW motion induced by SWs in AFMs compared to the same phenomena in FMs.

Spin-polarized currents can induce magnetization dynamics in magnetic materials~\cite{Brataas:2012fk}. However, DWs in FMs can also be moved by the transfer of spin angular momentum from travelling SWs, eliminating the additional dissipation cost associated with the electric current. Several theoretical ~\cite{Mikhailov:1984,*PhysRevLett.107.027205,*PhysRevLett.107.177207,*0295-5075-97-6-67002}, experimental~\cite{PhysRevLett.110.177202} and numerical~\cite{han:112502} studies have demonstrated that DW motion from magnonic spin transfer is possible. The reciprocal phenomenon has also been reported: DW motion in FMs induces local excitations of SWs ~\cite{PhysRevLett.65.2587,*PhysRevB.81.134405}.

\begin{figure}
\centering
\includegraphics[scale=0.44]{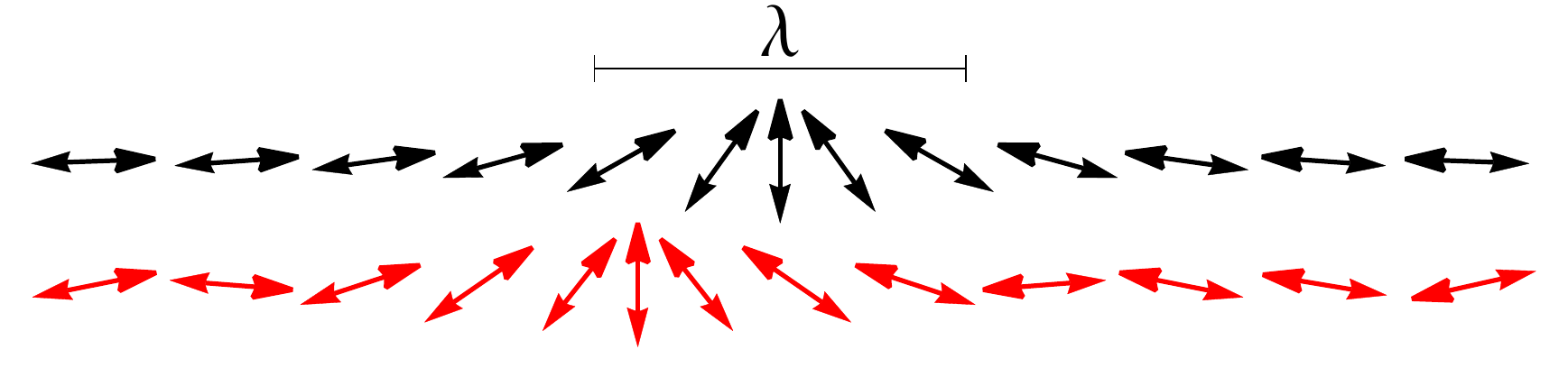}
\includegraphics[scale=0.44]{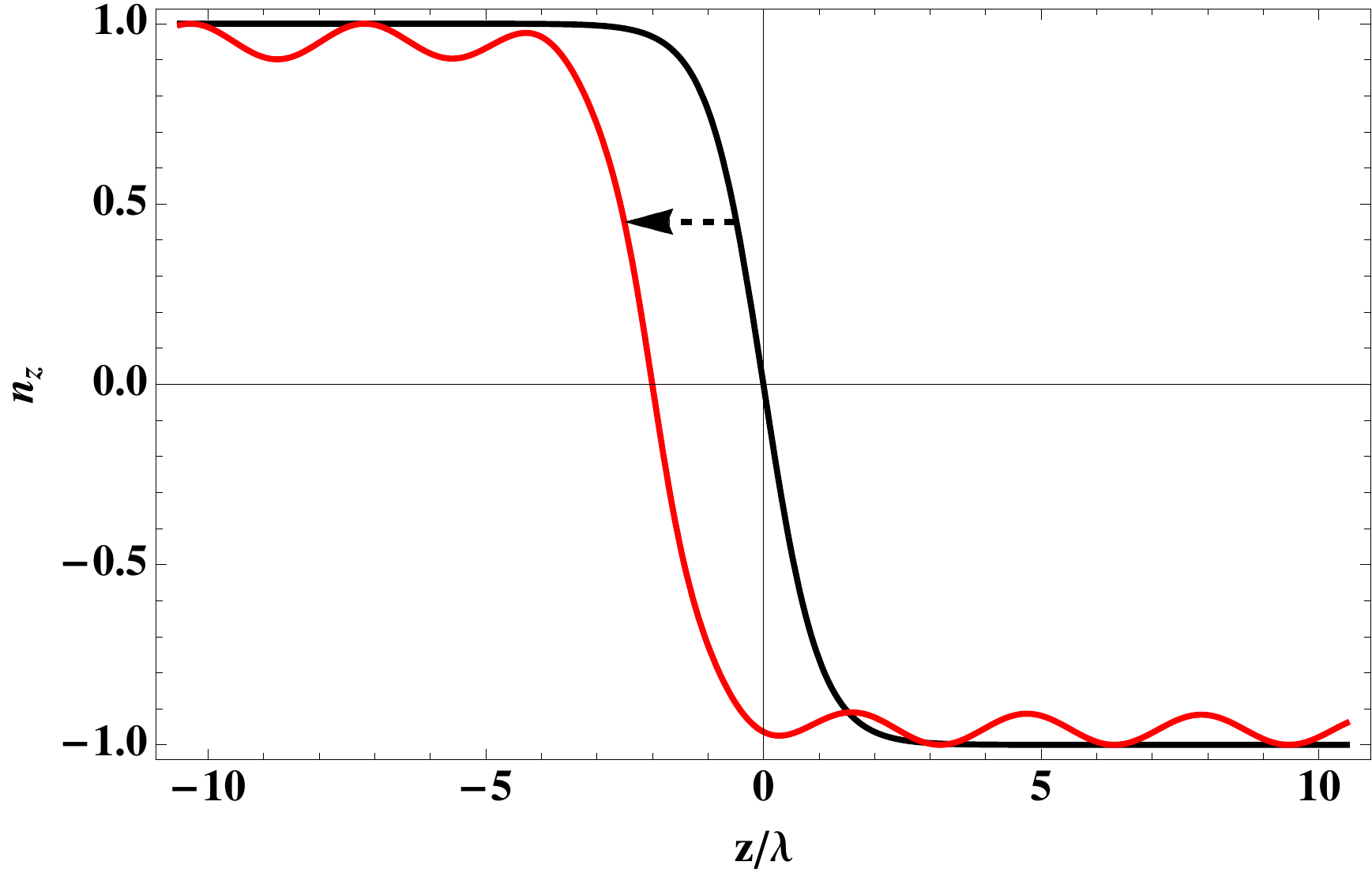}
\caption{(color online) Sketch of an antiferromagnetic DW displaced to the left as a result of linearly polarized SW excitations, where the SWs travel through the DW without reflection.}
\label{fig:DWdisplacement}
\end{figure}

In AFMs, circularly polarized SWs carry spin angular momentum whereas linearly polarized SWs do not. In a scenario in which a circularly polarized SW passes through an antiferromagnetic DW, the spin angular momentum flow associated with its motion is reversed. However, because the total spin angular momentum is conserved and antiferromagnetic DWs cannot absorb the constant transferred flux of spin angular momentum, we show that this scenario does not arise. Instead, circularly polarized SWs are reflected so that linear momentum is passed to the DWs. Here, we demonstrate that linearly polarized SWs, carrying no spin angular momentum, can pass through DWs without any reflection, as shown schematically in Fig.~\ref{fig:DWdisplacement}. As a result of this radical difference in the behavior of circularly and linearly polarized SWs, DWs move in opposite directions in response to the different modes of SW excitations.

The magnetizations on adjacent sublattices in an AFM are equal in magnitude but are oppositely directed. We consider a two-spin lattice, where the antiferromagnetic order parameter is defined as  $\vec{l}(\vec{r},t)=\vec{m}_{1}-\vec{m}_{2}$, and we introduce the normalized \textit{staggered} vector field $\vec{n}(\vec{r},t)=\vec{l}(\vec{r},t)/l$, where $l=|\vec{l}(\vec{r},t)|$. The total magnetization field $\vec{m}(\vec{r},t)=\vec{m}_{1}+\vec{m}_{2}$ is zero at equilibrium for AFMs. We also make use of the constraint $\vec{n}\cdot\vec{m}=0$, which is valid in the exchange approximation.

The equations of motion for the staggered field and the magnetization are as follows~\cite{PhysRevLett.106.107206}:
\begin{eqnarray}
\label{eq:LLn}
\dot{\vec{n}} & = &(\gamma \vec{f}_{m}-G_{1}\dot{\vec{m}})\times \vec{n}, \\
\label{eq:LLm}
\dot{\vec{m}} & = & (\gamma \vec{f}_{n}-G_{2}\dot{\vec{n}})\times\vec{n}+(\gamma \vec{f}_{m}-G_{1}\dot{\vec{m}})\times \vec{m},
\end{eqnarray}
where $\gamma$ is the gyromagnetic ratio, and $G_{1}$ and $G_{2}$ are phenomenological Gilbert damping constants. The effective fields $\vec{f}_{m}=-\delta_{\vec{m}} U$ and $\vec{f}_{n}=-\delta_{\vec{n}} U$ are functional derivatives of the free energy, $U$, of the AFM with respect to the magnetization and the staggered order, respectively.

We consider a one-dimensional texture, e.g., an insulating antiferromagnetic nanowire. In this case, the free energy is $U=\int d\vec{r}[a\vec{m}^2/2+A(\nabla \vec{n})^2/2-K_{z}n_{z}^2/2]$, where $a$ and $A$ are the homogeneous and inhomogeneous exchange constants, respectively. $a$ and $A$ are related through $a\sim A/(l^{2}d^{2})$~\cite{SovPhysUspBaryakhtar1985}, where $d$ is the lattice constant of the AFM. $K_{z}$ denotes the easy axis anisotropy along the wire, which is defined as the $z$ axis. We consider a DW created by pinning the AFM to adjacent FMs in different directions in the left and right reservoirs. The equilibrium shape of the DW is now determined by the competition between the exchange energy, $A$, and the anisotropy energy $K_{z}$.

To study the interaction between SWs and a DW, we perform a unitary transformation of the present coordinate system into the coordinate system of the DW, making use of the spherical unit vectors $\hat{r} = [\sin{\theta}\cos{\phi},\sin{\theta}\sin{\phi},\cos{\theta}]$, $\hat{\theta} = [\cos{\theta}\cos{\phi},\cos{\theta}\sin{\phi},-\sin{\theta}]$ and $\hat{\phi} = [-\sin{\phi},\cos{\phi},0]$. For a Walker DW~\cite{schryer:5406}, the equilibrium solution of Eqs.~\ref{eq:LLn} and \ref{eq:LLm} is given by $\vec{m}=0$ and $\theta_{0}=2\arctan\exp{(\xi)}$. $\xi$ is dependent on time through the DW center position, $r_{w}(t)$; $\xi=(z-r_{w})/\lambda$, where the DW width is defined as $\lambda=\sqrt{A/K_{z}}$. We also treat the out-of-plane angle $\phi_{w}(t)$ as a dynamic variable in the same manner as $r_{w}(t)$. 

SWs in AFMs are linear deviations of the staggered order, $\vec{n}(\xi,t)$, and the magnetization, $\vec{m}(\xi,t)$, around their equilibrium textures. The SW-DW interaction requires that we expand $\vec{n}(\xi,t)$ and $\vec{m}(\xi,t)$ 
to second order for small excitations, $h$, around the equilibrium DW texture $\hat{r}$:

\begin{eqnarray}
\label{eq:nexpansion}
\vec{n}(\xi,t) & = & \left[1-\frac{h^2}{2}(n_{\theta}^{2}(\xi,t)+n_{\phi}^{2}(\xi,t))\right]\hat{r}\nonumber\\
& &+h\left[n_{\theta}(\xi,t)\hat{\theta}+n_{\phi}(\xi,t)\hat{\phi}\right],\\
\label{eq:mexpansion}
\vec{m}(\xi,t) & = &h^2 m_{r}^{(2)}\hat{r}+\left[h m_{\theta}(\xi,t)+h^2 m_{\theta}^{(2)}(\xi,t)\right]\hat{\theta}\nonumber\\
& &+\left[h m_{\phi}(\xi,t)+h^2 m_{\phi}^{(2)}(\xi,t)\right]\hat{\phi},
\end{eqnarray}
where the notations $n_{\theta(\phi)}$ and $m_{\theta(\phi)}$ describe first-order excitations in the $\hat{\theta}(\hat{\phi})$ direction of the staggered field and the magnetization, respectively. We also include the second-order excitations in the magnetization, $m_{\theta(\phi)}^{(2)}$ and $m_{r}^{(2)}=-(m_{\theta}n_{\theta}+m_{\phi}n_{\phi})$.

Using the ansatzes Eqs.~\ref{eq:nexpansion} and \ref{eq:mexpansion} in Eqs. \ref{eq:LLn} and \ref{eq:LLm} and expanding the staggered field to the first order in $h$, we arrive at the equation of motion for SW excitations:
\begin{eqnarray}
\label{eq:spinwave}
\ddot{n}_{\theta(\phi)} & = & a K_{z}\gamma^2[\partial_{\xi}^2 n_{\theta(\phi)} +(2\mathrm{sech}^2(\xi)-1)n_{\theta(\phi)}]\nonumber\\
& &-a\gamma G_{2} \dot{n}_{\theta(\phi)} .
\end{eqnarray}

For simplicity, we have assumed that $G_{2}$ dominates $G_{1}$, simplifying the description of the SW dynamics. This assumption has been made only in the analytical treatment and is not included in the numerical results presented below. We conclude that excitations in the directions $\hat{\theta}$ and $\hat{\phi}$ are decoupled in AFMs, which is fundamentally different from the behavior of SWs in FMs~\cite{PhysRevLett.107.177207}. This result implies that both linearly and circularly polarized antiferromagnetic SWs exist.

Using $n_{\theta(\phi)}(\xi,t)=n_{\theta(\phi)}(\xi)\mathrm{exp}(-i\omega t)$, Eq.~\ref{eq:spinwave} reads 

\begin{equation}
\label{eq:schrodinger}
\hat{H}n_{\theta(\phi)}(\xi) = q^{2}n_{\theta(\phi)}(\xi),
\end{equation}
where the operator $\hat{H}=[-\partial_{\xi}^2-2\mathrm{sech}^2(\xi)]$. The eigenvalues $q^{2}=[\omega^2/(\gamma^2 a K_{z})-1+i\omega G_{2}/(\gamma K_{z})]$ define the dispersion relation of the antiferromagnetic SWs. Eq.~\ref{eq:schrodinger} is a time-independent Schr\"{o}dinger-type equation with the P\"{o}schl-Teller potential. This potential is reflectionless and offers exact solutions in the form of travelling wave eigenfunctions~\cite{:/content/aapt/journal/ajp/75/12/10.1119/1.2787015}. 

When $q$ is purely imaginary, the solutions to Eq.~\ref{eq:schrodinger}, $n_{\theta_{0}(\phi_{0})}=\rho_{0} \mathrm{sech}(\xi)$, where $\rho_{0}$ is an arbitrary amplitude, describe localized states, centered around the DW. These "Goldstone modes"~\cite{PhysRevB.79.174404} are distortions of the DW caused by the system being forced out of equilibrium and are naturally included in the formalism by considering the DW center, $r_{w}(t)$, and chirality, $\phi_{w} (t)$, to be collective dynamic variables of the system. 

For complex $q$, the solutions to Eq.~\ref{eq:schrodinger} represent propagating wave excitations superimposed on the staggered field texture. These solutions can be written as $n_{\theta(\phi)}(\xi, t)=\rho_{k}e^{i\Omega}(\tanh(\xi) - i q)$, where $\rho_{k}$ is the wavevector-dependent SW amplitude. $\Omega=q\xi-\omega t$, so that $\mathrm{Re}\{\Omega\}$ is the general phase of the wavelike excitations. Similar bound and travelling SW modes are also present in antiferromagnetic Bloch DWs~\cite{PhysRev.126.78}.

We use the ansatz that the accelerations of the DW center coordinate, $\ddot{r}_{w}(t)$, and chirality, $\ddot{\phi}_{w}(t)$, are proportional to the square of the amplitude of the SWs, and, thus, are second-order effects for the small excitation parameter, $h$. 

In the following, we assume that the antiferromagnetic SWs are linearly polarized transverse to the plane of the DW, along $\hat{\phi}$ ($n_{\theta}=0$ with $\phi(0)=0$). Circularly polarized SWs demand a different treatment and are discussed later. After combining Eqs.~\ref{eq:LLn} and \ref{eq:LLm}, inserting the effective fields  $\vec{f}_{n}=K_{z}n_{z}\hat{z}+A\partial_{z}^{2}\vec{n}$ and $\vec{f}_{m}=-a\vec{m}$, expanding to order $h^{2}$ and integrating over space, we find that $\ddot{\phi}_{w}+a\gamma G_{2}\dot{\phi}_{w}=0$, and the equation of motion for the DW coordinate, $r_{w}$, is
\begin{equation}
\label{eq:dwacc}
\ddot{r}_{w} + a\gamma G_{2}\dot{r}_{w}=\frac{a\gamma^{2}K_{z}}{\pi}\int_{-\infty}^{\infty}\mathrm{d\xi} \langle n_{\phi}^2\rangle\mathrm{sech}(\xi)\mathrm{tanh}(\xi),
\end{equation}
where $\langle n_{\phi}^2\rangle$ denotes a temporal average. By carrying out this average, we disregard temporal oscillations of the coordinate, $r_w$, as the DW moves.

Eq.~\ref{eq:dwacc} (without dissipation) is a result of the conservation of linear momentum density. As an explanation, let us consider the Lagrangian density of the AFM, $\mathcal{L}=\dot{\vec{n}}^2/(2a\gamma^{2})-A(\nabla\vec{n})^{2}/2+K_{z}n_{z}^{2}/2$~\cite{SovPhysUsp23.21,*PhysRevLett.50.1153}. Noether's theorem implies a continuity equation for the linear momentum density along $z$, $\mathrm{d}T_{zt}/\mathrm{dt}+\mathrm{d}T_{zz}/\mathrm{dz}=0$, where $T_{zj}(j=z,t)=(\partial_{z}q\ \partial_{\partial_{j} q}-\delta_{zj})\mathcal{L}$ is defined as in Ref.~\cite{PhysRevB.88.144413} and $q=\theta, \phi, n_{\phi}$. After integration and time averaging, we find that the continuity equation is identical to Eq.~\ref{eq:dwacc} (without dissipation).

The real part of the SW solutions for a small dissipation $G_{2}$ is 
\begin{equation}
\begin{split}
\label{eq:SWreal}
\mathrm{Re}\{n_{\phi}\}& \approx \frac{\rho_{k}}{(1+k^{2}\lambda^{2})^{1/2}}e^{-Q(\xi+|\xi_{0}|)/2} \\
& \hspace{-1.2cm}\times\left[\cos (k\lambda \xi-\omega t)\mathrm{tanh(\xi)}+k\lambda\sin (k\lambda \xi-\omega t)\right],
\end{split}
\end{equation}
 where $Q=G_{2}\omega/(\gamma K_{z} k \lambda)$, $\xi_{0}=(r_{w}-z_{0})/\lambda$, $z_{0}$ is the position of the excitation source, and $k=[\omega^{2}/(a\gamma^{2}K_{z})-1]^{1/2}/\lambda$ is the real wave vector of the monochromatic SWs at the driving frequency $\omega$. The SW amplitude depends on the form of the excitation source $H_{\mathrm{ext}}(z,t)$ through its spatial Fourier transform: $\rho_{k}=\omega\mathcal{F}_{k}\{H_{\mathrm{ext}}(z,t)\}/(a\gamma A k)$.

Inserting Eq.~\ref{eq:SWreal} into Eq.~\ref{eq:dwacc} and solving for the steady state ($\ddot{r}_{w}\rightarrow 0$) velocity, we obtain

\begin{equation}
\label{eq:velocity}
\dot{r}_{w}=-\rho_{k}^{2}e^{-Q|\xi_{0}|}\frac{(1+3k^{2}\lambda^{2})\omega}{6k}.
\end{equation}
Eq. \ref{eq:velocity}, which is our first central result, shows that the steady-state DW drift velocity induced by linearly polarized SWs is independent of the dissipation $G_{2}$ for high frequencies $\omega$. In the long-wavelength limit, when $k\lambda\rightarrow 0$, the DW velocity becomes large when the driving frequency is close to resonance, $\omega\rightarrow\omega_{0}=(a\gamma^2 K_{z})^{1/2}$. Naturally, the expansion in terms of a low dissipation $G_{2}$ breaks down close to this limit.

To verify Eq.~\ref{eq:velocity}, we conduct a micromagnetic simulation of Eqs.~\ref{eq:LLn} and \ref{eq:LLm} for a one-dimensional antiferromagnetic nanowire with a N\'{e}el DW in the $x$-$z$ plane as the initial condition. We add the external magnetic field source term $\vec{H}_{\mathrm{ext}}(z,t)$ to the free energy, $U\rightarrow U-\int \mathrm{d\vec{r}}\ \vec{H}_{\mathrm{ext}}\cdot\vec{m}$. We then write Eqs.~\ref{eq:LLn} and \ref{eq:LLm} in dimensionless form by scaling the time axis by $\tilde{t}=(\gamma a l)^{-1}$ and the $z$ axis by the lattice constant $d$. The simulation is based on the numerical method of lines with a time step control which is adaptive. The length of the wire is set to 1000 lattice constants, and the DW is initally positioned at $z=0$. We impose absorbing boundary conditions at $z\leq-400$ and $z\geq 400$. The SWs are excited in the region $z=[-72,-68]$ by a homogeneous and dimensionless magnetic field source $h_{\mathrm{ext}}(t)=\tilde{h}\sin{(\tilde{\omega}t)}\hat{x}$, where $\tilde{\omega}=\omega/\omega_{0}$. DW widths in AFMs are expected to be small~\cite{Bode:2006uq}, and therefore, we choose $\lambda=5d$. Other dimensionless constants are listed in Table \ref{tab:constants}.

\begin{figure}
\centering
\includegraphics[scale=0.85]{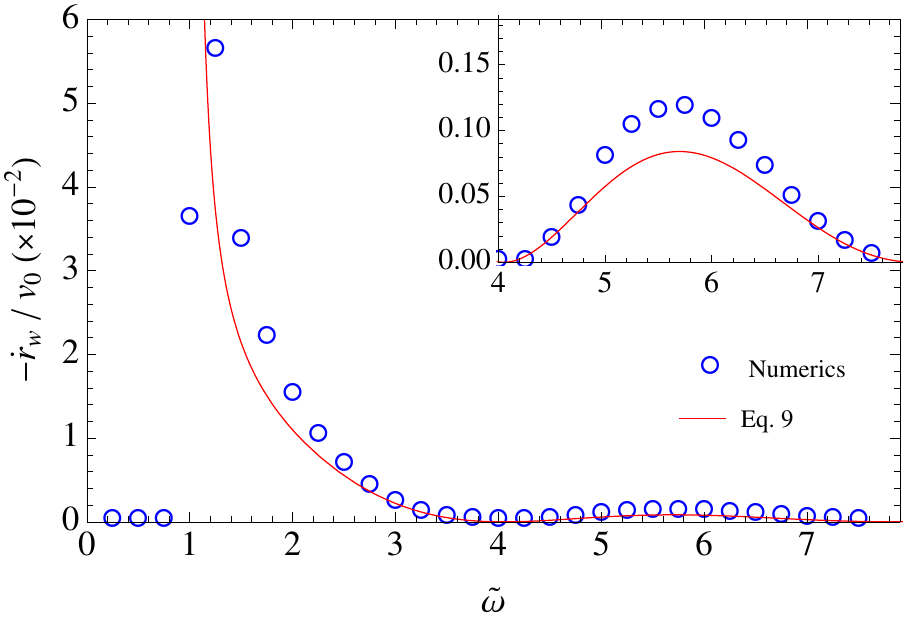}
\caption{(color online) Negative DW velocity, $-\dot{r}_{w}$, in units of $v_{0}=\gamma A/(l d)$, as a function of the applied excitation field frequency, $\tilde{\omega}$, for linearly polarized SWs. The DW is attracted towards the SW source. Blue circles represent the results of numerical simulations, and the red line indicates the analytical result based on Eq.~\ref{eq:velocity}.}
\label{fig:velvsfreq}
\end{figure}

Fig.~\ref{fig:velvsfreq} shows the simulated DW velocity as a function of excitation frequency. The velocity is given in units of $v_{0}=\gamma A/(l d)$. For frequencies close to $\omega_{0}$ the long-wavelength resonance peak is easily discernible. The velocity drops to zero for $\tilde{\omega}\approx 4$, which is a result of the step shape of the excitation source.

\begin{figure}
\centering
\includegraphics[scale=0.85]{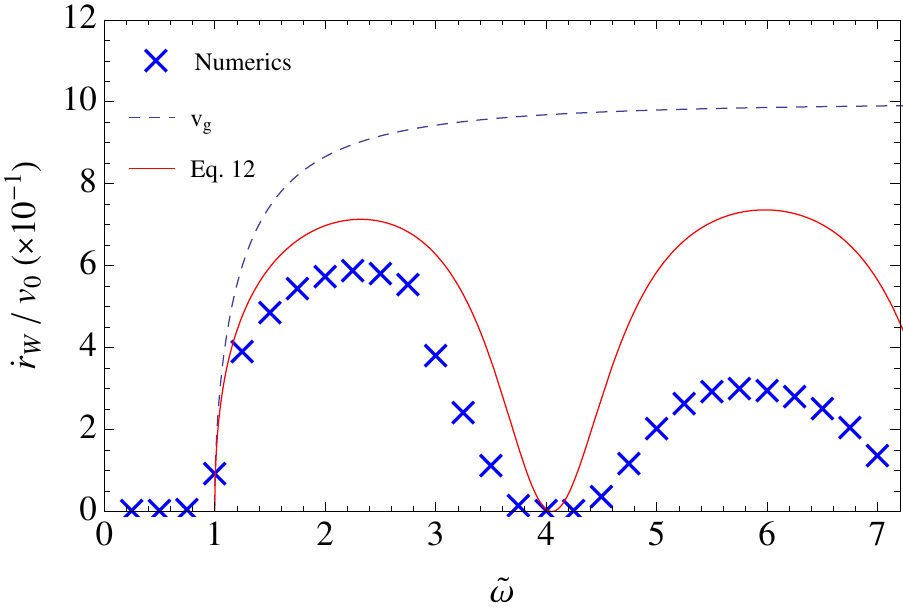}
\caption{(color online) DW velocity,  $\dot{r}_{w}$, as a function of the applied field frequency, $\tilde{\omega}$, for circularly polarized SWs. The DW is pushed away from the SW source due to reflection. Blue crosses represent the results from numerical simulations, the red line indicates the analytical result based on Eq.~\ref{eq:circvel}, and the dashed blue line shows the group velocity, $v_{g}$, of the SWs.}
\label{fig:velvsfreqcirc}
\end{figure}

\begin{table}[h!]
\centering
   \caption{Dimensionless numerical constants.}
   \begin{tabular}{@{} lcc @{}}
   \hline
      Constant          	& Composition               	&   Value             	\\
      \hline
      $\tilde{a}$           	& $a l^{2} d^{2}/A$     		&   1                   	\\
      $\tilde{G}_{1}$  	&   $G_{1}l$                    	& 0.002                 	\\
      $\tilde{G}_{2}$  	&   $G_{2}/l$                   	& 0.002                 	\\
      $\tilde{K}_{z}$ 		& $K_{z} d^2/A$             & $5^{-2}$                 	\\
      $\tilde{h}$ 				& $h/(a l)$          				& 0.05                 	\\ 
      $\tilde{z}_{0}$ 				& $z_{0}/d$               			& -70                 	\\
\hline
\end{tabular}
\label{tab:constants}
\end{table}

Although we consider an antiferromagnetic nanowire with easy axis anisotropy, we estimate the magnitude of the DW velocity using parameters for the antiferromagnetic insulator NiO, which has easy plane anisotropy in the bulk. We use $A_{\mathrm{NiO}}\approx 5\times 10^{-13\ }$ J/m, $d_{\mathrm{NiO}}\approx 4.2\ \mathrm{\AA}$, and a magnetic moment per sublattice of $1.7\mu_{B}$~\cite{PhysRevB.84.115114}, with $\mu_{B}$ being the Bohr magneton. With these parameters $v_{0}\approx 500$ m/s, and the resonance frequency $\omega_{0}\approx 200$ GHz. The DW drift velocity induced by long-wavelength linearly polarized SWs in $\mathrm{NiO}$ is then approximately 5-10 m/s directed toward the SW source.

Next, we discuss the very different interactions that arise between circularly polarized SWs and DWs. Numerically, when we excite circularly polarized antiferromagnetic SWs, $h_{\mathrm{ext}}(t)=\tilde{h}[\sin{(\tilde{\omega} t)}\hat{x}+\cos{(\tilde{\omega} t)}\hat{y}]$, we observe that the SWs are reflected from the DW structure (SW behavior not shown). The DW now moves in the same direction as the incoming SWs. This DW behavior is opposite to that observed for linearly polarized SWs. Additionally, circularly polarized SWs also cause the DW to acquire a net angular velocity, $\dot{\phi}_{w}$.

To elucidate this phenomenon, consider Eq.~\ref{eq:LLm} with the effective fields inserted but without dissipation:
\begin{equation}
\label{eq:magnetizationconservation}
\dot{\vec{m}}=-\gamma\vec{n}\times K_{z} n_{z}\hat{z}-\nabla \vec{J}_{m},
\end{equation}
where $\vec{J}_{m}=\gamma A\vec{n}\times\nabla\vec{n}$ is defined as the \textit{spin wave spin-current}~\cite{Kajiwara:2010fk} through the AFM. The $z$ component of Eq.~\ref{eq:magnetizationconservation} has the form of a conservation law for spin angular momentum, $\partial_{t}m_{z}+\partial_{z}J_{m_{z}}=0$, where $J_{m_{z}}(\xi)=\gamma A (n_{\theta}\partial_\xi n_{\phi}-n_{\phi}\partial_\xi n_{\theta})\mathrm{tanh}(\xi)/\lambda$. The spin wave spin-current vanishes for linearly polarized SWs, whereas circularly polarized SWs carry $J_{m_{z}}=\pm \gamma A k \rho_{k}^{2}$, where the sign depends on the SW helicity. After integration over space, we find
\begin{eqnarray}
\label{eq:magnz}
\partial_t M_{z} & = & -\left[ J_{m_{z}}(\infty)-J_{m_{z}}(-\infty)\right],
\end{eqnarray}
where $M_{z}$ is the total magnetization in the $z$ direction.

There are two possibilities for circularly polarized SWs. In the first scenario, the SWs are transmitted through the DW, causing the spin-current to change its sign after transmission. In this case, the right-hand side of Eq.~\ref{eq:magnz} is finite, which leads to the build-up of a local magnetic moment around the DW. In the second scenario, the SWs are reflected, and the right-hand side of Eq.~\ref{eq:magnz} vanishes. Only the second scenario is possible in the steady state because the strong exchange interaction in the AFM counteracts the build-up of an increasing local magnetic moment.

Having established that circularly polarized SWs are reflected, we calculate the DW velocity by means of linear momentum transfer from reflected SW packets to the DW. From the Lagrangian density, we calculate the linear momentum density in the $z$ direction, $T_{zt}=\partial_{z}q\partial_{\dot{q} }\mathcal{L}$~\cite{PhysRevB.88.144413}, with $q=\theta,\phi,n_{\theta},n_{\phi}$. After integrating over space, we find that the total linear momentum in the $z$ direction, $P_{z}=\int\mathrm{dz}\ T_{zt}$, can be split into a DW part and a SW part: $P_{z}^{\mathrm{DW}}=2\dot{r}/(a\gamma^2\lambda)$ and $P_{z}^{\mathrm{SW}}=\int \mathrm{dz}(n_{\theta}^{2}+n_{\phi}^{2})k \omega/(a\gamma^{2})$. When considering SW packets, the continuity equation for linear momentum density in the $z$ direction becomes a conservation law for the total linear momentum, $P_{z}$, according to Noether's theorem, and we find $0=\mathrm{d}P_{z}/{\mathrm{dt}}=\mathrm{d}/\mathrm{dt}(P_{z}^{\mathrm{DW}}+P_{z}^{\mathrm{SW}})$. A train of reflected SW packets exerts a force $\Delta P_{z}^{\mathrm{SW}}/\Delta t = \rho_{k}^{2}k\omega (v_{g}-\dot{r}_{w})/(a\gamma^{2})$, where $v_{g}= a\gamma^{2}A k/\omega$ is the SW group velocity. Balancing this force to the force on the DW, $\mathrm{d}P_{z}^{\mathrm{DW}}/{\mathrm{dt}}=2(\ddot{r}_{w}+a\gamma^2 G_{2}\dot{r}_{w})/(a\gamma^2\lambda)$, gives the resulting DW velocity in steady state as
\begin{equation}
\label{eq:circvel}
\dot{r}_{w}=\frac{v_{g}}{1+\frac{a\gamma G_{2}}{\rho_{k}^{2}\lambda k\omega}},
\end{equation}
which is our second and final central result. For low damping, the DW is accelerated to the SW group velocity, $v_{g}$, which is several hundred meters per second for typical AFMs. Therefore, the DW motion induced by circularly polarized SWs is oppositely directed and much faster than the motion caused by linearly polarized SWs. Numerically, we see in Fig.~\ref{fig:velvsfreqcirc} that Eq.~\ref{eq:circvel} captures the DW motion well, especially at low applied frequencies. We speculate that the assumption of total SW reflection will break down for higher frequencies. 

In conclusion, we have investigated the manner in which antiferromagnetic SWs move DWs in AFMs. Linearly polarized SWs drive DWs towards the SW source, analogous to the effect of magnon spin transfer torque in FMs. In contrast to the ferromagnetic case, where the DW moves due to the conservation of angular momentum, the SW-driven antiferromagnetic DW motion can be understood as arising from the conservation of linear momentum density. Circularly polarized antiferromagnetic SWs are scattered by the DW to prevent the build-up of a local magnetic moment around the DW center. This behavior causes the DW to move away from the SW source at velocities of several hundred meters per second.

\end{document}